\documentclass[runningheads]{llncs}

\usepackage{graphicx}
\newcommand{\exl}{{ExLibris }}
\newcommand{\exlt}{{ExLibris}}
\newcommand{\opt}[1]{{\tt #1}}
\newcommand{\lib}[1]{{\tt #1}}
\newcommand{\prd}[2]{{\tt #1}/\oldstylenums{#2}}
\newcommand{\prdthr}[3]{{\tt #1}/\oldstylenums{#2},\oldstylenums{#3}}
\newcommand{\trm}[1]{{\tt #1}}
\newcommand{\var}[1]{\ensuremath{#1}}
\newcommand{\atm}[1]{{\tt #1}}
\newcommand{\file}[1]{{\tt #1}}
\newcommand{\xop}[2]{{\tt #1}(\ensuremath{#2})}

\begin{document}
\setcounter{page}{89}
\title{Exporting Prolog source code}
\titlerunning{Exporting Prolog source code}
\author{Nicos Angelopoulos}
\authorrunning{N. Angelopoulos}
\institute{Department of Computing,\\
           Imperial College, London.\\
           \email{nicos@doc.ic.ac.uk}}

\maketitle

\addtocounter{footnote}{1}
\footnotetext{In Alexandre Tessier (Ed), proceedings of the 12th International Workshop on Logic Programming Environments (WLPE 2002), July 2002, Copenhagen, Denmark.\\Proceedings of WLPE 2002: \texttt{http://xxx.lanl.gov/html/cs/0207052} (CoRR)}

\begin{abstract} 
In this paper we present a simple source code configuration
tool. ExLibris operates on libraries and can be used to extract
from local libraries all code relevant to a particular project.
Our approach is not designed to address problems arising in
code production lines, but rather, to support
the needs of individual or small teams of researchers
who wish to communicate their Prolog programs.
In the process, we also wish to accommodate and encourage
the writing of reusable code. Moreover, we support and propose ways
of dealing with issues arising in the development of code that
can be run on a variety of \emph{like-minded} Prolog systems.
With consideration to these aims we have made the following decisions:
(i) support file-based source development, 
(ii) require minimal program transformation,
(iii) target simplicity of usage, and 
(iv) introduce minimum number of new primitives.
\end{abstract} 

\section{Introduction}

Prolog has been around for nearly thirty years.
Its ability to survive as a general purpose programming
language can be mainly attributed to the fact that 
it is complimentary to the major players in the field.
Without disregard to the many commercial products written
in Prolog, the language, arguably, thrives in
academic environments, and in particular in AI and 
proof-of-concept computer science research. 

An important element in such projects is that the
developers are only expected to write code in a 
part-time basis within a volatile environment.
As a result, programs evolve from
few hundred lines to several thousands in an evolutionary
manner, that is, without prior overall design of the 
final product. Indeed, it is seldom the case that 
an identifiable final product stage is ever reached.

This is contrary to expectations in non-academic settings.
As is the fact that sharing and publishing of unfinished
source code is desirable. Furthermore tools such as 
the Unix {\tt make} utility, \cite{Feldman79}
which admittedly targets a different set of objectives,
requires duplication of work and discourages re-usability
of Prolog code. In contrast, we present
\exl which  makes use of the directives present in
Prolog source files to overcome these problems.

A convenient method for including relatively positioned
source code is by using the \opt{library} alias present
in most modern Prolog systems. This mechanism is 
used primarily for system code that implements 
useful common predicates. For example the \lib{lists}
library present in most Prolog systems
defines, among others, predicates 
\prd{member}{2} and \prd{append}{3}.
\exl extends the idea by allowing, during project development,
access to code from a number of \emph{home} library directories.
When one
wants to export the project for public use, the
source files that are relevant are bundled into a local
library directory. The only change required is that 
the local directory is added as a library directory in 
the top source files.

This library oriented approach encourages the writing of 
reusable code. For instance, predicates that accomplish generic
tasks should be developed in the home library.
Furthermore, it promotes a library oriented way of thinking,
where useful code can become independent and in later
stages part of the system libraries.
For example, the Pillow program
\cite{Cabeza97} has been incorporated in the current
SICStus 3.9.0 release \cite{Sicstus:3:9:0}.

Unlike the $DERIVE$ system, \cite{Brereton95}
we have chosen to use the underlying file store, and to 
provide in-source support for system-dependent configuration.
$DERIVE$ stores predicates in a relational database
and uses table attributes to achieve a more holistic
approach to Prolog based software engineering.

Dependence on source files means that in order to accommodate
multiple prolog engines and runtime loading we need to 
introduce some new primitives. In this paper we present
a minimum set of such primitives which we believe are 
interesting in, at least, pointing some of the support needed for
such tasks.

\exl can be used for configuring both coarse and fine
grain libraries. Coarse libraries define many predicates
per file, whereas finer grains reduce this to a possibly
minimum of one predicate per file. \exl depends for the grouping 
of source files to the
primitives provided by the file-system, that is on the 
subdirectory relation.

The remaining of this paper is organised as follows.
Section 2 deals with some preliminary Prolog definitions
that deal with conditional loading and tentative dependencies
of source files. Section 3, presents the functionality of \exlt.
Section 4, provides some comments on the features, limitations,
and possible future work. Finally, Section 5 serves as the
concluding section.

\section{Preliminaries}

\begin{figure}[tb]
        \begin{center}
                \includegraphics*[height=0.363\textheight]{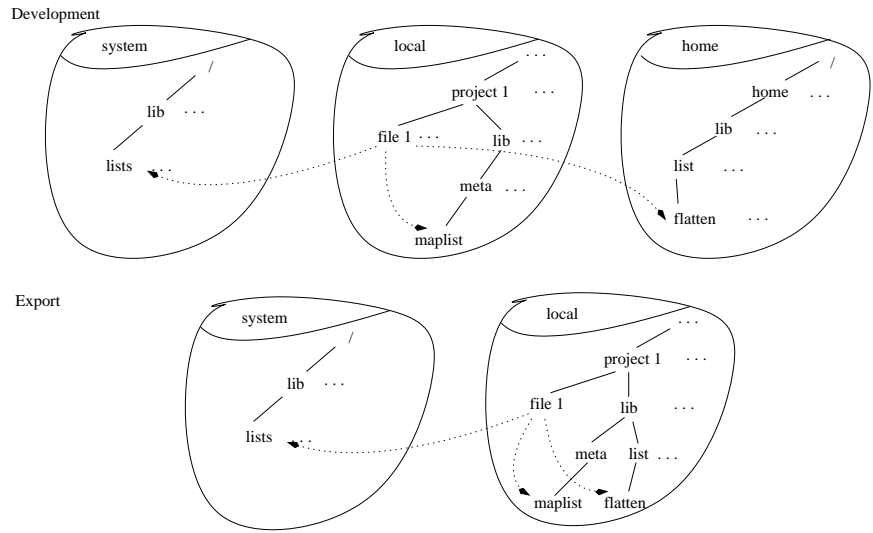}
        \end{center}
        \caption{all sorts of libraries}
        \label{fig:libs}
\end{figure}

The standard development and configuration phases supported
by \exl are shown in Fig.~\ref{fig:libs}. Development happens
at a project directory which, possibly, contains a local
library directory. During this phase (an example depicted in top
part of Fig.~\ref{fig:libs}) files in the project
directory can use the \opt{library} alias to load any
of the following
three: system files; which are part of the supported
prolog engines, home files; part of the developer's
or the developing team's filespace, and local files; which
are within the project's space. In top part of Fig.~\ref{fig:libs}
\file{file1} of \file{project1} depends on files
\file{lists}, \file{maplist} and \file{flatten} which reside
within system, local and home libraries respectively.

\exl is a tool that helps 
to create an export directory that is independent 
of home library dependencies. This is illustrated in the 
lower part of Fig.~\ref{fig:libs}. In the exported code,
dependencies are either to system or to local libraries.
All files that are relevant to files in \file{project1}
and reside within the home library are copied across
to the local directory. Although single system, local and home
directories are shown in Fig.~\ref{fig:libs}, multiple, or alternative
as is the case for system, ones, are supported.

\subsection{Conditional load predicates}
Since the publication of the Prolog ISO standard \cite{Iso95,Deransart96}
the vast majority of systems have strove for compliance.
This has made the idea of the same prolog code running
on different engine feasible.
Still differences exist,
and it is necessary to take these into account.

The two issues we need to address are, 
uniform structured prolog identification and 
conditional loading. These tasks are useful in
their own right, so we collect the relevant predicates
in the pl library. This has been implemented and
tested for SICStus, SWI \cite{Swi:5:0:5}, and Yap \cite{Yap:4:3:20}.

\subsubsection{pl/1}
Firstly, \lib{pl} defines predicate \prd{pl}{1}(\trm{pl(-Pl)}) .
Its argument identifies
the running prolog system with a compound term, we 
refer to this term as \trm{pl-term}. The name 
of the functor of the term identifies the prolog system and the term's
single argument the version, \trm{pl-version}. The version should be such
that the term order imposes the relevant order on the 
versions.
For example the terms for the
three most recent versions of the supported systems are:
sicstus(3:9:0), swi(5:0:7), and yap(4:3:23).

\subsubsection{defines/1,2}
\label{sec:defines}
A first use for \prd{pl}{1} is for defining suitability
of particular prolog engines for predicates present in a source file.
We define directives,
\begin{verbatim}
:- defines( +Functors ).
:- defines( +IfPls, +Functors ).
\end{verbatim}
We will treat the first directive as a special case of the second
where \var{IfPls} is instantiated to \atm{all} indicating suitability
for all Prolog systems. Apart from the special value \atm{all}
\var{IfPls} may be a single, or a list of \trm{if-pl-term}. In turn,
an \trm{if-pl-term} may be a \trm{if-term} or a pair of the form
(\var{PlName},\var{PlConds}). \var{PlName} is a valid name for the functor
of some \trm{pl-term} and \var{PlConds} is a list of 
(\var{PlVers},\var{PlOper}) pairs.

The intuition is that definitions for predicates corresponding to
\var{Functors} are given in the source file in which the directive appear,
provided
that the underlying engine matches one of the \trm{if-pl-terms} in
\var{IfPl}. To find a match, elements in \var{IfPl} are considered
disjunctively whereas \var{PlConds} are considered conjunctively.

\subsubsection{mkindex/1}
Predicate \prd{mkindex}{1} takes a number of options that are
not discussed here in full. Its core operation is to create 
an index file (\file{Index.pl}) for a given, library, directory.
The index file, contains fact definitions for \prd{index}{5}
multifile predicate.
\begin{verbatim}
index( -Name, -Arity, -IfPls, -Module, -File ).
\end{verbatim}
\var{Name} and {Arity} refer to the functor of a predicate defined
in file \var{File} and module \var{Module}. \var{IfPls} is of the 
same form as that described in Section~\ref{sec:defines} and
indicate that the definition in \var{File} should only be used for
Prolog system with matching \trm{pl-term}.

\prd{mkindex}{1} has three sources for finding this information.
Firstly, module definitions (in which case \var{IfPls} == \atm{any})
the \prdthr{defines}{1}{2} directives as already defined and in the case
where neither of the two is present, it can be instructed to attempt
and extract it from the clausal definitions. In the last case
\var{IfPls} is again instantiated to \atm{any}.

\subsubsection{requires/1}
With indices such as those described above,
source files can ensure the presense of specific predicates
without reference to the executing Prolog system
while ensuring that the correct versions will be loaded at 
loading time. 
\begin{verbatim}
:- requires( +Functors ).
\end{verbatim}
Directive \prd{requires}{1} instructs that at loading time
the files corresponding to each functor and the running
system will be loaded. 

Following from the example in Fig.~\ref{fig:libs} at 
development phase the mentioned files might contain
\begin{verbatim}
\% file1
:- requires( [member/2,maplist/3,flatten/2] ).

\% SysLib/Index.pl
index( member, 2, sicstus(_), lists, lists ).

\% LocalLib/Index.pl
index( maplist, 3, any, user, 'meta/maplist' ).

\% HomeLib/Index.pl
index( flatten, 2, swi(_), built_in, 'compat/swi/built_ins' ).
index( flatten, 2, not(swi(_)), user, 'list/flatten' ).
index( member, 2, swi(_), built_in, 'compat/swi/built_ins' ).
\end{verbatim}

Module name \trm{built\_in} means that 
the predicate is a built-in.
As can be seen above, \prd{flatten}{2} is a built-in in Swi
so no file will be loaded. The file present is
simply where our instruction that this is a built-in
for this system resides. On the other hand if \file{file1}
is loaded in SICStus then the definition in 
\file{HomeLib/list/flatten} will be used.
Similarly for \prd{member}{2} although in the 
later case it will be loaded from \file{SysLib/lists}.
For either system \prd{maplist}{3} is loaded from 
\file{LocalLib/meta/maplist}.

\subsubsection{if\_pl/2,3}
\label{sec:plload}
Files can also be loaded conditionally to the 
current system by using the introduced \trm{pl-term} within
the loading file.
Predicates \prdthr{if\_pl}{2}{3} provide
means for accomplishing this, and can be called as follows:
\begin{verbatim}
if_pl( +IfPls, +Call ).
if_pl( +IfPls, +Call, +ElseCall ).
\end{verbatim}
The predicate is quite general since calls \var{Call} and
\var{ElseCall}
can be any callable term. Here we are interested in 
the cases where \trm{if\_pl} is used as a directive and
the calls is of the form $LoadCall(\ldots Files \ldots)$.
\var{IfPls} is identical to the one  described
in Section~\ref{sec:defines}. Here, if the running 
engine has a match, then \var{Call} is called. In \prd{if\_pl}{3}
\var{ElseCall} is called if there is no match.

\exl recursively de-constructs \prd{if\_pl}{2,3} directives,
and recognises any loading predicates within \var{Call} and
\var{ElseCall}, and thus can provide information similar to
to the that gathered in \prd{index}{5} by \prd{mkindex}{1}.
Contrary to \prd{index/5} the information is present on 
the caller file rather than the defining file, so it 
is not as clean an approach. We have included \prd{if\_pl}{2,3}
firstly because it is a useful predicate taking advantage
of \prd{pl}{1} and secondly because it might be useful
in cases when one chooses not to use \file{Index.pl} files.

\subsection{Dependent files}
Finally, we need to address a discrepancy that 
arises from loading code at runtime.
Unlike when using
directives these situations give no easily accessible 
information about
the files a program depends upon.  Although it seems
useful to have a directive declaring tentative dependencies
such feature is not present in any of the discussed 
systems.

We propose a very simple mechanism facilitated by
\prd{may\_load}{1} directives. :- may\_load( +Files ).
declares that a single or a list of files may be loaded
at runtime by the program present in the same source file.

In this way we make the dependency of the two files 
more accessible. The dependant file is the file in which
the directive appears in. This, depends on the file
pointed to by the argument of \prd{may\_load}{1}. 
Such a directive assists \exl in ensuring that all files
that may be loaded, depending on the execution path of 
a predicate, will be exported.

\section{Export}
Predicate \prd{exlibris}{1} is used to create an export
directory structure from the developer's source code.
The emphasis is placed in
integrating relevant parts of private libraries when building
software for exportation.
Its single argument is a list of options. The recognised
options are as follows.
\begin{description}
        \item{\xop{dest}{Destination}} the destination directory
                where the exported files will be copied. This should not 
                exist prior to the call. 
                This option does not have a default value.

        \item{\xop{source}{Srcs}} a single file or directory or a list of
                source files and directories. Each is considered to be either
                an \emph{entry} level source file, or a directory containing
                entry level source files.
                An entry level file is one that a user is expected to load 
                directly. In the case of directories all 
                source files within are considered entry level source files.
                There is no default value for this option.

        \item{\xop{copy}{Copy}} whether directories containing entry
                files should also be copied recursively, \var{Copy} == recursive,
                or entry files should be copied individually, \var{Copy} == selective.
                Default is \var{Copy} == selective.
        
        \item{\xop{syslib}{SysLib}} usually is a single system library path, but 
                a list of paths can also be given. The provided  path should point to 
                the developing Prolog's system library directory. Default is the
                first directory given as the answer to query
                ?- library\_directory( L ). and conforms to system dependant
                criteria. Effectively we try to guess which library directory
                is the system one, since this is not an information Prolog
                engines currently provide.

        \item{\xop{homelibs}{HomeLibs}} a list of private libraries holding
                source files that are loaded from entry files or their 
                dependents by the \opt{library} alias. The intuition is that 
                during development
                these directories are defined using\\ \prd{library\_directory}{1}
                in entry files or some appropriate start file. The default value is 
                the instantiations of ?- library\_directory( L ). that didnt
                match the \atm{syslib} criteria.

        \item{\xop{loclib}{LocLib}} a path for the local library.
                This is considered relatively to \var{Destination}. 
                All referenced files in \var{HomeLibs} will be copied into
                $LocLib$. The relative path of any such file from the 
                appropriate \var{HomeLib} will be recreated within \var{LocLib}.
                Note that this may be an existing directory within some source 
                directory.  Default value: \atm{lib}.

        \item{\xop{pls}{Pls}} a single or a list of \trm{pl-terms}.
                Only files pertinent to systems corresponding
                to these \trm{pl-terms} are copied. These are identified
                from \prd{if\_pl}{2,3} directives as discussed in 
                Section~\ref{sec:plload}. The default value is for 
                all prologs which is equivalent to \xop{pls}{all}

\end{description}

The exported files are identical to the development ones
proviso two transformations. Entry level files loose any
\prd{library\_directory}{1} definition and instead the
following lines are added on the top of each such file
\begin{verbatim}
% Following line added by ExLibris.
:- library_directory( 'RelPathToLocLib' ).
\end{verbatim}
When exporting, the value of directory \var{RelPathToLocLib}
is known and it is the path to \var{LocLib} relative to the
particular entry level file.
The second transformation is to remove any \prd{if\_pl}{2,3} that 
does not match any of the system \trm{pl-term} in \var{Pls}.

\section{Discussion}

Our approach uses the file-system's directory structure
as its medium of grouping predicates at the level 
of source files. This, supports both fine and coarse 
grain groupings. Examples of coarse groupings are 
the system libraries defining a score of predicates for
source file. Whereas, fine grouping would favour 
single predicate definition per source file or module
files exporting a single predicate.
However, operations such as
moving source files within the home directory structure
will need to be accommodated by future tools.

Currently, \exl runs on SICStus v3.9.0. Swi v5.0.7 and
Yap v4.3.23 (cvs) or later under Unix-like file-systems.
Our belief is that like-minded systems such as
Ciao \cite{Ciao:7:1} and GnuProlog \cite{GnuProlog:1:2:13}
will be easy to support.
For SICStus and Yap, and since  \opt{layout} option only provides 
the start line of read terms, \exl requires that 
\trm{if\_pl} terms are the only terms on the text line in which 
they appear, and also that there are no new line
characters to the end of the term (to the period).
Other operating systems may be supported via the
support Prolog systems provide for translation of Unix paths
to other operating system paths.
All code described in this paper can be found
at http://www.doc.ic.ac.uk/\verb+~+nicos/exlibris/ .
About half of the code used was code drawn from pre-existing
private libraries.

A number of additional tools may be constructed that
can help with keeping projects and libraries consistent
as well as facilitating library merging. For such tasks,
as is also true for other source code manipulation,
it will be useful to have a structured form of comments.

In the future we will like to implement non-recursive 
library copies. That is, the relative path of a home library
file is reconstructed into the exported local library
directory.
This feature is currently not supported because it requires 
code transformations to a degree greater than we wish the
core program to have.
One possibility would be to add this as an additional 
tool that can flatten out any arbitrary library while
updating project source files and inter-library dependencies.

\section{Conclusions}
The first contribution of this paper was to propose
simple mechanisms for conditional, depending on the 
underlying system, loading and execution,
and for declaring tentative source file dependencies.
Apart from the suitability of the particular suggestions
it is important that some of the issues raised here and which
have remained dormant should attract some attention from the Prolog
community.

Based on the proposed primitives we also presented a straightforward
procedure for code configuration and exportation. We have kept core 
functionalities to a minimum as to encourage simplicity
and thus usage. 
\exl encourages development of reusable code. We perceive this
as the most desirable feature of \exlt.


\newcommand{\etalchar}[1]{$^{#1}$}

\end{document}